\documentclass[twocolumn,notitlepage,prx,superscriptaddress,longbibliography]{revtex4-2}
\usepackage{amsfonts}
\usepackage{textcomp}
\usepackage{times}
\usepackage{graphicx}
\usepackage{float}
\usepackage{latexsym,amsmath,amssymb,bm,euscript}
\usepackage{color}
\usepackage{subfigure}
\usepackage{epstopdf}
\usepackage[colorlinks=true,linkcolor=blue,citecolor=blue]{hyperref}
\usepackage{hyperref}
\usepackage{soul}
\usepackage[normalem]{ulem}
\usepackage{mathrsfs}
\usepackage{amsmath}
\usepackage{xspace}
\usepackage{natbib}
\usepackage{ulem}
\usepackage{etoolbox}
\usepackage{tikz}

\newcommand{\Eq}[1]{Eq.~\eqref{#1}}

\newcommand{\Fig}[1]{Fig.~\ref{#1}}
\newcommand{\Figs}[1]{Figs.~\ref{#1}}

\begin{document}

\title{Phase Diagram, $d$-Wave Superconductivity, and Pseudogap \\
of the $t$-$t'$-$J$ Model at Finite Temperature}

\author{Dai-Wei Qu}
\affiliation{Kavli Institute for Theoretical Sciences, University of Chinese Academy of Sciences, Beijing 100190, China}
\affiliation{Institute of Theoretical Physics, Chinese Academy of Sciences, Beijing 100190, China}

\author{Qiaoyi Li}
\affiliation{Institute of Theoretical Physics, Chinese Academy of Sciences, Beijing 100190, China}
\affiliation{School of Physical Sciences, University of Chinese Academy of Sciences, Beijing 100049, China}

\author{Shou-Shu Gong}
\affiliation{School of Physical Sciences, Great Bay University, Dongguan 523000, China}
\affiliation{Great Bay Institute for Advanced Study, Dongguan 523000, China}

\author{Yang Qi}
\email{qiyang@fudan.edu.cn}
\affiliation{Department of Physics, Fudan University, Shanghai 200433, China}
\affiliation{Hefei National Laboratory, Hefei 230088, China}

\author{Wei Li}
\email{w.li@itp.ac.cn}
\affiliation{Institute of Theoretical Physics, 
Chinese Academy of Sciences, Beijing 100190, China}
\affiliation{Hefei National Laboratory, Hefei 230088, China}

\author{Gang Su}
\email{gsu@ucas.ac.cn}
\affiliation{Kavli Institute for Theoretical Sciences, University of Chinese 
Academy of Sciences, Beijing 100190, China}
\affiliation{Institute of Theoretical Physics, 
Chinese Academy of Sciences, Beijing 100190, China}

\begin{abstract}
Recently, robust $d$-wave superconductive (SC) order has been unveiled in the 
ground state of the 2D $t$-$t'$-$J$ model --- with both nearest-neighbor ($t$) and 
next-nearest-neighbor ($t'$) hoppings --- by density matrix renormalization group 
studies. However, there is currently a debate on whether the $d$-wave SC holds 
up strong on both $t'/t>0$ and $t'/t<0$ cases for the $t$-$t'$-$J$ model, which 
correspond to the electron- and hole-doped sides of the cuprate phase diagram, 
respectively. Here we exploit state-of-the-art thermal tensor network approach to 
accurately obtain the phase diagram of the $t$-$t'$-$J$ model on cylinders with 
widths up to $W=6$ and down to low temperature as $T/J \simeq 0.06$, pushing 
the boundaries of contemporary finite-$T$ calculations. For $t'/t>0$, we find a 
domelike SC regime with a diverging $d$-wave pairing susceptibility, $\chi_\textrm{SC} 
\propto 1/T^\alpha$ below a characteristic temperature $T_c^*$. Near optimal doping, 
$T_c^*$ reaches its highest value of about $0.15 J$. Above $T_c^*$ yet below a higher 
crossover temperature $T^*$, the magnetic susceptibility becomes suppressed, 
which can be related to the onset of pseudogap (PG) behaviors. On the other hand, 
for $t'/t<0$ we find the pairing correlations are much weaker, although there exhibits 
a node-antinode structure in the PG regime as observed in the hole-doped cuprates. 
The thermal tensor network calculations of the $t$-$t'$-$J$ model underscore both 
the similarities and differences in the finite-temperature phase diagram between the 
fundamental model and cuprates, yielding unique insights into their intricate 
behaviors.
\end{abstract}
\maketitle

\textit{Introduction.---}
Understanding unconventional superconductive (SC) phase and the enigmatic normal states 
like those in the pseudogap (PG) regime of the cuprate phase diagram has become one of 
the major challenges in modern condensed matter physics~\cite{Cuprate1986,TsueiRMP2000,
LNW-HighTc-RMP2006,Keimer2015Nature,Proust2018Review,Arovas2022Review}. Theoretically, 
the two-dimensional (2D) Hubbard~\cite{Hubbard1963a,Gutzwiller1963a} and $t$-$J$ models
\cite{ZhangRiceSinglet1988,Spalek2007} are believed to capture the essence of electron 
correlations~\cite{Anderson1987RVB,Anderson2004}. Despite active and intensive studies
\cite{White1998tJ,Huscroft2001DCA,White2003Stripes,Georges1996RMP,Maier2005RMP,
Kotliar2006RMP,Assaad2008,Gu2010Grassmann,Chang2010,Chan2012DMET,Gull2013SCPG,Corboz2014tJ,
LeBlanc2015PRX,Fradkin2015RMP,Wu2017DiagramQMC,Dodaro2017tJ,Zheng2017Science,
He2019AFQMC,Corboz2019Stripe,HCJiang2018tJ,HCJiang2019Science,HCJiang2020PRR,Chung2020Plaquette,
Qin2020Absence,HCJiang2021tJ,Gong2021tJ,Jiang2021PNAS,Qin2022Review,Xu2022Stripes,Dong2022NatPhys,
Dong2022PNAS,Xiao2023OrdersPRX,Jiang2023tJ8leg,Lu2024tJ8leg,Jiang2024Hubbard6leg,Xu2024Science}, it remains 
elusive whether these fundamental models can reproduce the $d$-wave SC and PG states 
observed in the phase diagram.

Recently, great efforts have been devoted to the accurate calculations of Hubbard and $t$-$J$ models, 
significantly advancing their understanding. It has been shown that in the Hubbard model with only
nearest-neighbor (NN) hopping and a large repulsive $U$, long-range SC order is absent under $1/8$ 
hole doping while a stripe order appears instead~\cite{White1998tJ,White2003Stripes,Corboz2014tJ,
Zheng2017Science,Corboz2019Stripe,Qin2020Absence,Xu2022Stripes}. Lately, by considering the 
next NN hopping $t'$, large-scale density matrix renormalization group (DMRG) studies revealed 
suppression of the stripe order and the emergence of robust $d$-wave SC
\cite{HCJiang2021tJ,Gong2021tJ,Jiang2021PNAS,Jiang2023tJ8leg,
Lu2024tJ8leg,Jiang2023tJ8leg,Jiang2024Hubbard6leg,Xu2024Science}.

Given the discoveries at zero temperature, it is natural to inquire whether the $t$-$t'$-$J$ model also 
contains exotic many-electron states at elevated temperatures. There are compelling questions to be 
answered, including whether the superconducting ground state expands to a SC phase at finite temperature; 
if yes, what is the characteristic temperature $T_c^*$? Does this imply the existence of high-temperature SC 
in the 2D limit? Are there exotic normal states akin to the PG states present? To address these intriguing 
questions, an efficient and unbiased finite-temperature approach is essential. 

While the finite-temperature Lanczos method is 
limited within small system sizes~\cite{FTLanczos,Prelov1996tJ,Jaklic2000Review,Prelov2013} 
and the quantum Monte Carlo approaches~\cite{Assaad2008} suffer from the notorious 
sign problem at finite doping, the thermal tensor network (TN) provides a powerful framework to 
simulate correlated systems at large scale, which are witnessing a rapid development in recent 
years~\cite{White2009METTS,Stoudenmire2010,Wietek2021PRX,Wietek2021TLU,Li2011a,
Czarnik2014PEPO,Czarnik2016PRB,Czarnik2019PRB,Czarnik2019PRBII,Aritra2022Hubbard,
Chen2018XTRG,HLi2019PRB,Chen2019TLH,Li2020NC,HLi2020PRR,Li2021NC,YuCPL2021,
Chen2022tbg}. In particular, the tangent space tensor renormalization group ($\tan$TRG)
\cite{tanTRG2023} effectively bridges the gap between ground-state and finite-temperature TN 
calculations. Akin to DMRG, its computational cost scales as $O(D^3)$, where $D$ is the bond 
dimension controlling the computational accuracy. 

In this work, we study the $t$-$t'$-$J$ model on width-4 and -6 cylinders using the state-of-the-art 
thermal TN approach~\cite{Chen2018XTRG,Chen2021SLU,tanTRG2023}. We have established 
the doping-temperature phase diagram containing both $t'/t>0$ and $t'/t<0$ cases, and reveal that 
the $d$-wave SC states exist in a domelike regime on the $t'/t>0$ side (electron doping), where the 
SC susceptibility $\chi_\textrm{SC}$ exhibits a power-law divergence. The estimated SC temperature is as
high as $T_c^* \simeq 0.15 J$ near the optimal doping, confirming the robust $d$-wave SC order 
observed in the ground state, which indeed corresponds to high-temperature SC. Above the SC phase, 
we find the magnetic susceptibility exhibits a maximum at around $T^*$ that represents the crossover 
temperature between the PG and high-temperature (HT) regimes. $T^*$ decreases with increasing 
the doping level, resembling the temperature scale of the PG regime observed in the doped cuprates
\cite{Johnston1989,Alloul1989KightShift,Monien1991PRB}. On the other hand, on the $t'/t<0$ side 
(hole doping) we find vanishingly small pairing correlation. Nevertheless, calculations of the Matsubara 
Green's function unveil node-antinode structures in the Fermi surface within the PG regime, evident 
on both $t'/t > 0$ and $t'/t < 0$ sides.
Besides, the antiferromagnetism (AFM) and charge density wave (CDW) instabilities are also discussed. 
Our results reveal a domelike $d$-wave SC phase with high $T_c^*$ in a basic $t$-$t'$-$J$ 
model, clarifying key aspects of its finite-temperature phase diagram and underscoring both the similarities 
and distinctions with cuprate superconductors.

\textit{Model and methods.---}
The Hamiltonian of the square-lattice $t$-$J$ model reads
\begin{equation}
H = - \sum_{i,j,\sigma} t_{ij} (c_{i\sigma}^\dagger c_{j\sigma}^{\,} + \textrm{H.c.})
    + \sum_{i,j} J_{ij} (\mathbf{S}_i \cdot \mathbf{S}_j - \frac{1}{4}n_i n_j ),
\label{Eq:Ham}
\end{equation}
where $c_{i\sigma}^\dagger$ ($c_{i\sigma}^{\,}$) is the electron creation (annihilation) 
operator with spin $\sigma = \uparrow, \downarrow$, $\mathbf{S}_i$ denotes the 
spin-$1/2$ operator, and $n_i = n_{i\uparrow} + n_{i\downarrow}$ is the particle number operator. 
The double occupancy in the local Hilbert space is projected out. We consider the NN 
hopping $t$, the next-nearest-neighbor (NNN) hopping $t'$, and also the NNN $J'$ as $J/J'=(t/t')^2$. $J\equiv1$ 
is set as the energy scale, and the NN hopping is fixed as $t/J=3$ (implying $U/t =12$ 
for the original Hubbard model). The implementation of U(1)$_\textrm{charge} \times$ 
SU(2)$_\textrm{spin}$ symmetry with QSpace~\cite{Weichselbaum2012,Weichselbaum2020,
Weichselbaum2024} significantly enhances the calculation efficiency.

The geometries explored include width-4 cylinders (e.g., $4\times30$ with width $W=4$ 
and length $L=30$), and width-6 systems (e.g., $6\times18$ cylinder and $6\times6$ open 
square). On long cylinders, the SC state takes on a form of a Luther-Emery liquid with 
both quasi-long-range CDW and $d$-wave SC correlations, where the spin and single-particle 
excitations are gapped~\cite{Balents1996,White2002Friedel,HCJiang2018tJ,HCJiang2019Science,
Chung2020Plaquette,HCJiang2020PRR,Gannot2020Ladder,HCJiang2020Critical}. 
Below, we consider the two cases of $t'/t=\pm0.17$ [with $J'/J =(t'/t)^2\approx 0.03$] 
and explore the corresponding phase diagrams. 

In the thermal TN simulations, we take the grand canonical ensemble with a 
chemical potential term $-\mu\sum_i n_i$ controlling the hole doping. We retain 
U(1)$_\textrm{charge} \times $ SU(2)$_\textrm{spin}$ multiplets for the width-4 (up 
to $D^*=4000$ multiplets, about $12,000$ individual states) and width-6 systems 
(up to $D^*=6000$ multiplets, about $21,000$ states), and obtain well converged
results~\cite{Supplementary}. For 
computing the magnetic (SC pairing) susceptibilities, a small magnetic (pairing) 
field is applied, where the U(1)$_\textrm{charge} \times$ U(1)$_\textrm{spin}$ 
[$\mathbb{Z}_{2, \textrm{charge}}$ $\times$ SU(2)$_\textrm{spin}$] symmetry 
is exploited. We have also conducted DMRG calculations as references 
in the zero-temperature limit, by retaining 4096 U(1)$_\textrm{charge} \times$ 
SU(2)$_\textrm{spin}$ multiplets (about $11,000$ individual states), where the 
small truncation error of $\epsilon \lesssim 10^{-6}$ ensures high accuracy.

% ================ Fig. 1 ================= %
\begin{figure}[!tbp]
\includegraphics[width=1\linewidth]{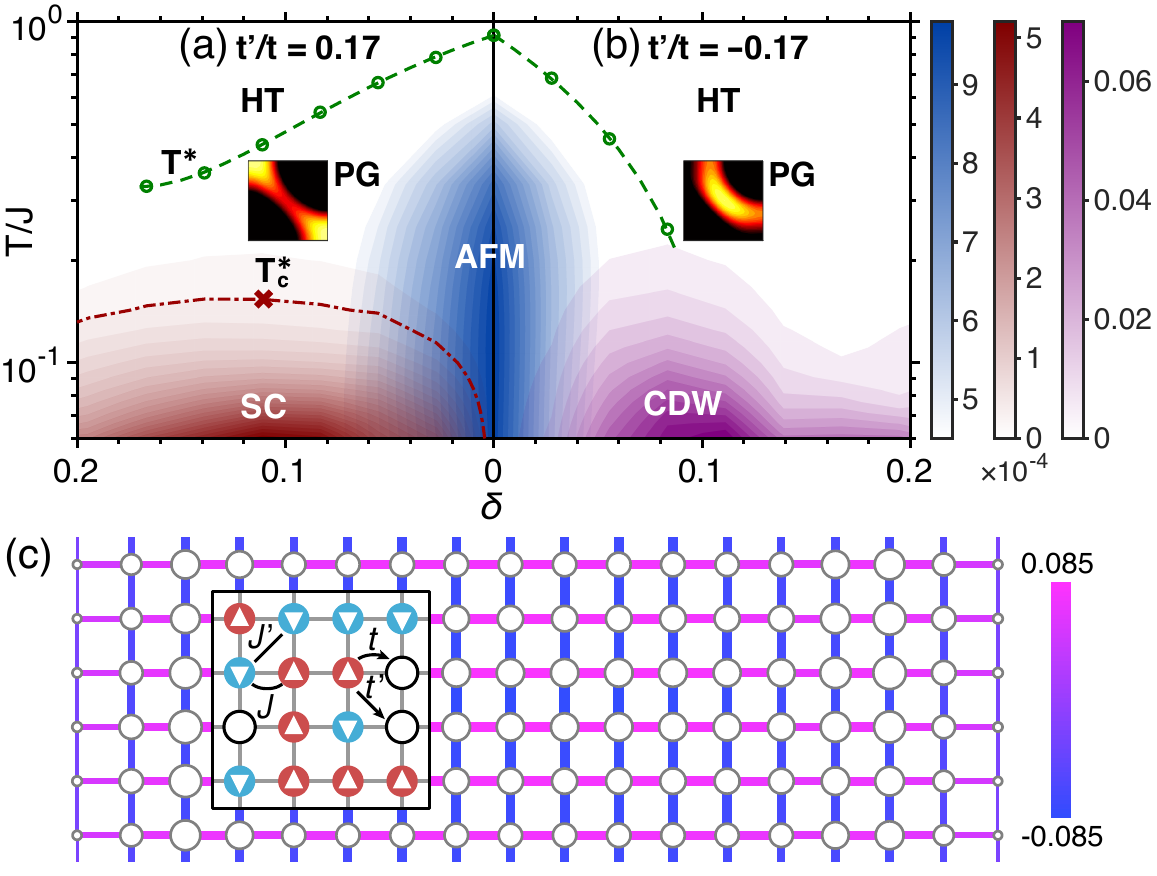}
\caption{Finite-temperature phase diagram of the $t$-$t'$-$J$ model for 
(a) $t'/t=0.17$ and (b) $t'/t=-0.17$, on the width-6 cylinder. The red contour 
shows $\Phi_{yy}(r)$ values averaged over $r=3$-$5$ along the cylinder, which forms a domelike 
regime with strong SC fluctuations. The dotted dashed line denotes $T_c^*$ 
determined from pairing susceptibility [cf. Fig.~\ref{Fig:Chi}(a)]. In (b), 
the pairing strength is negligibly small for $t'/t = -0.17$. The blue contour illustrates 
AFM spin structure $S(\pi,\pi)$, and the purple CDW regime emphasizes the 
electron-density peak $\max{(\rho_k)}$. The green dashed line indicates the 
crossover $T^*$ from the HT to PG regimes, marked by a suppression in 
magnetic susceptibility $\chi_m$. The insets display the spectral density on 
the Fermi surface for both $t'/t=0.17$ and $-0.17$ in the PG regime (cf., 
Fig.~\ref{Fig:Matsubara}). (c) illustrates the $d$-wave pairing on a $6 \times 18$ 
cylinder at $T/J = 1/14$, for $t'/t = 0.17$ and doping $\delta \simeq 0.1$. 
Bond thickness and color denote pairing strength and sign, while circle size indicates 
hole density. The inset depicts the $t$-$t'$-$J$ model, with empty circles representing 
holes and red (blue) circles indicating spin-up (spin-down) electrons.}
\label{Fig:PhDiag}
\end{figure}

\begin{figure}[!tbp]
\includegraphics[width=1\linewidth]{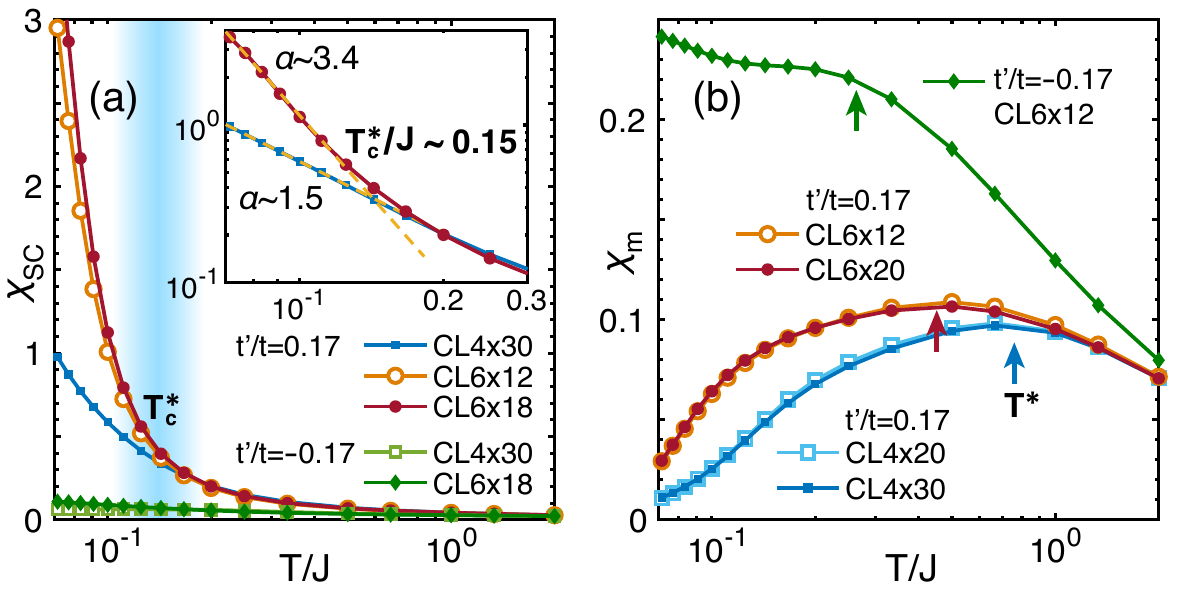}
\caption{(a) $d$-wave pairing susceptibility $\chi_\textrm{SC}$, 
where the inset shows the power-law divergence of $T^{-\alpha}$ 
below $T_c^*\sim 0.15J$. For the $W=4$ data, the chemical potential 
is fixed at $\mu=5.34$ for $t'/t=0.17$ and $\mu=5.45$ for $t'/t=-0.17$; 
for all $W=6$ data we fix $\mu=5.1$. They render doping levels varying 
from $\delta \simeq 0.15$ to $\delta\simeq 0.1$ as temperature lowers. 
(b) shows the magnetic susceptibility $\chi_m$, where the hump or shoulder 
at $T^*$ is indicated by the arrow, with doping fixed at $\delta \simeq 0.1$. 
The results are obtained with up to 3000 $\mathbb{Z}_{2, \textrm{charge}}$ 
$\times$ SU(2)$_\textrm{spin}$ multiplets in (a) and 
8000 U(1)$_\textrm{charge} \times$ U(1)$_\textrm{spin}$ states in (b).
}
\label{Fig:Chi}
\end{figure}

\textit{Doping-temperature phase diagram.---} 
We first demonstrate our main findings for $t'/t>0$ in the phase diagram 
Fig.~\ref{Fig:PhDiag}(a), where $\delta = (1-n)$ measures the hole doping 
(with $n$ the electron density). We characterize SC by computing the pairing
correlation $\Phi_{\alpha\beta}(\mathbf{r}) = \langle 
\Delta_\alpha^\dag(\mathbf{r}_0) \Delta_\beta^{\,}(\mathbf{r}_0 
+ \mathbf{r}) \rangle$ with $\Delta_\alpha(\mathbf{r}) = \frac{1}{\sqrt{2}} 
(c_{\mathbf{r},\downarrow} c_{\mathbf{r} + \alpha,\uparrow} - 
c_{\mathbf{r},\uparrow}c_{\mathbf{r} +\alpha,\downarrow})$ being the (singlet) 
pairing field operator ($\alpha$ is the unit vector $\hat{x}$ or $\hat{y}$). 
In Fig.~\ref{Fig:PhDiag}(a), we display the color contour of the pairing correlations 
$\Phi_{yy}$, where a domelike regime extending from the SC ground state is 
identified. The pronounced SC fluctuations within this domelike regime support 
the robust SC order reported in previous DMRG  studies~\cite{HCJiang2021tJ,
Gong2021tJ,Jiang2021PNAS,Jiang2023tJ8leg}. As shown in Fig.~\ref{Fig:Chi}(a), there exists 
a characteristic SC temperature scale $T^*_c \sim 0.15 J$ (i.e., $\sim 0.05 t$), 
below which the pairing susceptibility $\chi_\textrm{SC}(T)$ falls into an algebraic 
divergence versus $T$. 

Moreover, we compute the singlet-pairing density matrix $\rho_S(\mathbf{r}_i, 
\alpha | \mathbf{r}_j, \beta) = \langle \Delta_\alpha^\dag(\mathbf{r}_i) 
\Delta_\beta(\mathbf{r}_j) \rangle$, whose dominant eigenvector can 
provide insight into the SC order and pairing pattern~\cite{Wietek2022,
Baldelli2023Fragmented}. To exclude local contributions, the matrix 
element is set as 0 if $\Delta_\alpha^\dag(\mathbf{r}_i)$ and 
$\Delta_\beta(\mathbf{r}_j)$ overlap. In Fig.~\ref{Fig:PhDiag}(c) we show 
the dominant eigenvector computed on a $6\times 18$ cylinder system 
with doping $\delta\simeq 0.1$. A robust $d$-wave pattern with uniform
charge distribution appears at low temperature $T/J=1/14$, fully consistent 
with previous DMRG calculations~\cite{Gong2021tJ}. 

Above the SC dome, we find another temperature scale $T^*$ derived from 
the magnetic susceptibility $\chi_m$ shown in Fig.~\ref{Fig:Chi}(b). The location 
of the $\chi_m$ peak naturally defines $T^*$, and the intermediate-temperature 
regime between $T_c^*$ and $T^*$ resembles the PG regime observed in the 
cuprate phase diagram. Above the PG regime there exists a HT regime, 
which is possibly a metallic phase influenced by many-electron correlations 
yet deformed by the no-double-occupancy constraint in the $t$-$J$ model. 

To explore the magnetic correlations, we compute the spin structure factor $S(\mathbf{k}) 
= \frac{1}{N}\sum_{i,j} e^{\mathrm{i} \mathbf{k}\cdot (\mathbf{r}_i-\mathbf{r}_j)} \langle 
\mathbf{S}_i\cdot \mathbf{S}_j \rangle$, where $N=WL$. In Fig.~\ref{Fig:PhDiag}(a), we present the contour 
plot of $S(\pi,\pi)$, which reveals a region with pronounced AFM correlations near half filling 
(see also Supplemental Material~\cite{Supplementary}). At half filling, the $t$-$J$ model 
reduces into the Heisenberg model, which exhibits a single peak in its specific heat at 
$T_\textrm{AFM}/J\simeq 0.6$~\cite{HLi2019PRB}. Below this temperature, AFM correlations 
rapidly strengthen. We use this as the criterion for determining the AFM regime, adjusting 
the color limits to ensure that the top of the contour aligns with $T_\textrm{AFM}/J \simeq 0.6$, 
and find that the regime rapidly diminishes with increasing doping.  

For the $t'/t<0$ case, on the other hand, we have done similar analysis, and the results are 
summarized in Fig.~\ref{Fig:PhDiag}(b). In contrast to the $t'/t>0$ case, the pairing correlations 
are significantly weaker, even negligible, different from experimental findings in hole-doped 
cuprates. The AFM regime is more constrained than its counterpart in the $t'/t>0$ side, which 
aligns with experiments. Although the SC order is found negligible for $t'/t<0$, we observe 
prominent CDW modulations at low temperature~\cite{Supplementary}. In Fig.~\ref{Fig:PhDiag}(b), 
we present a contour plot of $\max(\rho_k)$, which reflects the CDW modulation (charge stripe) 
with momentum $k$, i.e., $\rho_k = \frac{1}{\sqrt{L}} \sum_x e^{\mathrm{i}k x}[n(x)-n]$. 
Additionally, under certain doping we observe a $\pi$-phase shift in the spin correlation,
which is indicative of spin stripes~\cite{Supplementary}, and these spin stripes precede the 
formation of charge stripes, or CDWs, as temperature decreases.

% ======= Fig. 2 ====== %

% ======= Fig. 3 ======== %
\begin{figure}[!tbp]
\includegraphics[width=1\linewidth]{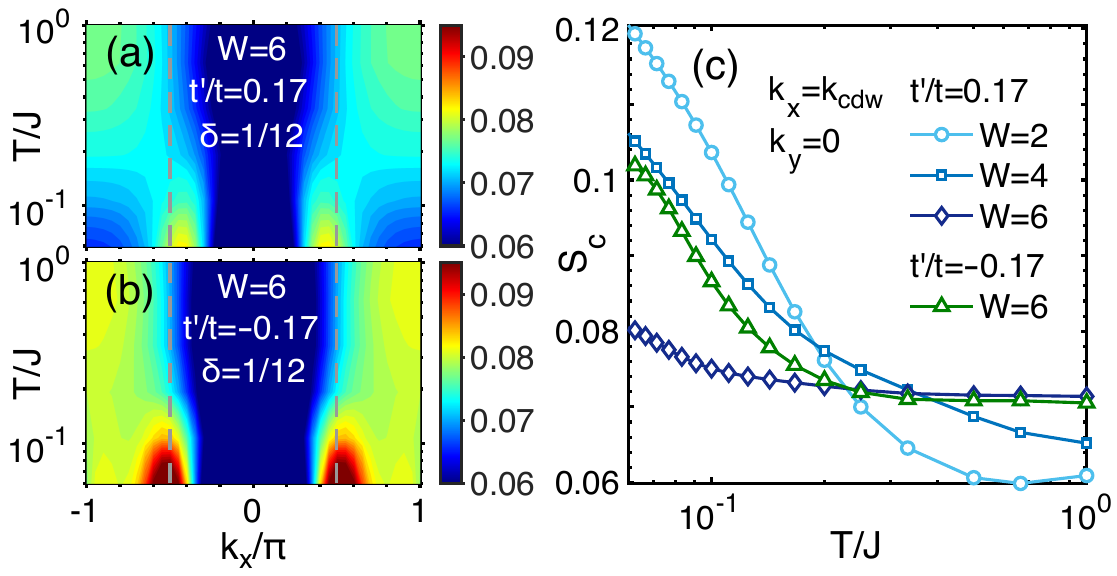}
\caption{(a),(b) The contours of charge structure factor $S_c(k_x,k_y=0)$, 
for $t'/t=0.17$ and $-0.17$, respectively. The results are calculated on $W=6$ cylinders. 
The dashed gray lines indicate the CDW wave vectors $k_\textrm{CDW}=\pm6\delta\pi$. 
(c) shows the temperature dependence of $S_c(\mathbf{k}_\textrm{CDW})$ at the corresponding 
CDW wave vectors for different $W$ and $t'/t$. The doping is approximately fixed 
at $\delta=1/12$ by fine-tuning the chemical potential $\mu$.
}
\label{Fig:SF}
\end{figure}

\textit{Pairing and magnetic susceptibilities.---}
To understand the two temperature scales $T_c^*$ and 
$T^*$, we compute the pairing and magnetic susceptibilities. The $d$-wave 
pairing susceptibility is defined as $\chi_\textrm{SC} = (1/N) \, \partial \langle 
\Delta_\textrm{tot} \rangle / \partial h_\textrm{p}$, which is computed by applying 
a pairing field term $-h_\textrm{p} \Delta_\textrm{tot}$, with $\Delta_\textrm{tot} 
= \sum_{\mathbf{r},\alpha} f_\alpha [\Delta_\alpha(\mathbf{r}) + 
\Delta_\alpha^\dag(\mathbf{r})]/2$ ($f_{\hat{x}} = -1$ and $f_{\hat{y}} = 1$) and a 
small field $h_\textrm{p}=0.01$. In Fig.~\ref{Fig:Chi}(a), for $t'/t=0.17$ we find $\chi_\textrm{SC}$ 
diverges algebraically below $T_c^*$ (see the log-log plot in the inset). Upon increasing 
the system width from 4 to 6, we find that the $\chi_\textrm{SC}$ results almost coincide 
on $W=4$ and $6$ cylinders for $T>T_c^*$; for $T<T_c^*$, the $W=6$ curve diverges 
much faster with a significantly larger exponent $\alpha$. We anticipate that  
$\chi_\textrm{SC}$ for wider systems will continue to coincide for $T>T_c^*$, 
and diverges even more rapidly for $T \lesssim T_c^*$. $T_c^*$ thus provides 
a good upper bound for the true $T_c$ in the 2D limit. Moreover, the specific heat 
results also support this conclusion, where the strong pairing fluctuations contribute 
to a prominent specific heat peak near $T_c^*$~\cite{Supplementary}. In contrast, 
as also depicted in Fig.~\ref{Fig:Chi}(a), only very small pairing susceptibility $\chi_\textrm{SC}$ 
can be observed for $t'/t=-0.17$, which, in conjunction with the weak pairing correlations 
indicated in Fig.~\ref{Fig:PhDiag}(b), suggests the absence of SC phase for typical 
dopings on $W=6$ cylinder with $t'/t<0$. 

In \Fig{Fig:Chi}(b), we show the magnetic susceptibility 
$\chi_m$ for $t'/t=0.17$, computed under a small pinning magnetic field of $0.01$. 
We find it exhibits a maximum at $T^*$, indicating 
the onset of PG~\cite{Huscroft2001DCA,Rubtsov2009PRB,Gull2013SCPG}. 
Moreover, $\chi_m$ decreases as $T$ lowers, reflecting 
a finite spin gap in the SC state. In Fig.~\ref{Fig:PhDiag}(a), we show that $T^*$ 
lowers as the doping level increases. As doping ratio $\delta$ exceeds a 
certain value, $\chi_m$ does not vanish as $T \rightarrow 0$, suggesting 
that the spin gap closes in the overdoped regime. 
For $t'/t=-0.17$, while the magnetic susceptibility $\chi_m$ also displays 
a hump at $T^*$, it keeps increasing as the temperature lowers, as illustrated
in Fig.~\ref{Fig:Chi}(b) and also in Supplemental Material~\cite{Supplementary}. 
This constitutes another striking difference between the $t'/t>0$ and $t'/t<0$ sides, 
and offers additional evidence for the absence of SC for the latter case. In Fig.~\ref{Fig:PhDiag}(b), 
$T^*$ decreases more sharply with doping for 
$t'/t=-0.17$ compared to the $t'/t=0.17$ case in Fig.~\ref{Fig:PhDiag}(a), 
aligning with the observation of narrower AFM regime for $t'/t=-0.17$. 

\textit{CDW correlations.---}
In Fig.~\ref{Fig:SF}, we show the computed charge structure factor $S_c(\textbf{k}) = \frac{1}{N}
\sum_{i,j} e^{\textrm{i}\textbf{k} \cdot(\mathbf{r}_i-\mathbf{r}_j)}\langle 
(n_i-n)(n_j-n) \rangle$, where $n=1-\delta$ is the electron density. 
The doping ratio $\delta \simeq 1/12$ is selected by tuning the chemical 
potential $\mu$. In Fig.~\ref{Fig:SF}(a) we find, for $t'/t=0.17$ and $W=6$, 
that the CDW peaks at  $k_x\simeq\pm \pi/2$ emerge for $T \lesssim T_c^*$. 
The CDW peaks $k_x\simeq\pm W\pi\delta$ also emerge for $W=2,4$, 
and the charge distribution $n(x)$ oscillates in real space below $T_c^*$, 
shaking hands with the ground-state DMRG results~\cite{Supplementary}. 
For the case of $t'/t=-0.17$, as shown in Fig.~\ref{Fig:SF}(b) the 
$S_c(\textbf{k})$ also exhibits CDW peaks at $k_x\simeq\pm \pi/2$ below 
a certain temperature, which are much more prominent than those for 
$t'/t=0.17$. In Fig.~\ref{Fig:SF}(c), we show $S_c(\mathbf{k}_\textrm{CDW})$ 
versus $T$ and find for $t'/t=0.17$ that it gets significantly suppressed as 
width $W$ increases, consistent with the conclusion in previous DMRG 
studies~\cite{HCJiang2020PRR,Lu2022tJPRB,Gong2021tJ}.
Conversely, for $t'/t=-0.17$, $S_c(\mathbf{k}_\textrm{CDW})$ remains significant
for $W=6$, suggesting stronger CDW fluctuations in the low-temperature regime 
with a typical doping of $\delta=1/12$. This scenario was also observed in recent 
DMRG studies \cite{Jiang2021PNAS,Jiang2024Hubbard6leg,Lu2024tJ8leg}. 
For $t'/t<0$, $W=4$ is special by hosting a distinctive plaquette $d$-wave 
ground state~\cite{Chung2020Plaquette}, and thus its charge structure factors 
are not shown.

\textit{Fermi surface topology.---}
The suppression of magnetic susceptibility below $T^*$ suggests the presence 
of PG behaviors. To validate it, we calculate the Matsubara Green's function 
$G(\mathbf{k}, \beta/2) = \sum_\sigma \langle e^{\beta H/2} c^\dag_{\mathbf{k}\sigma} 
e^{-\beta H/2} c_{\mathbf{k}\sigma} \rangle_\beta$, where $c_{\mathbf{k}\sigma} = \frac{1}{\sqrt{N}}
\sum_\mathbf{r} e^{-\mathrm{i}\mathbf{k}\cdot\mathbf{r}} c_{\mathbf{r}\sigma}$. 
This imaginary-time proxy provides estimate of the 
spectral weight around the Fermi level as $\beta G(\mathbf{k}, \beta/2) \sim A(\mathbf{k},\omega=0)$ 
in the low-temperature limit~\cite{ZYMeng2022NC,Lederer2017PNAS}, 
thus reflecting the Fermi surface (FS) topology. 

\begin{figure}[!tbp]
\includegraphics[width=1\linewidth]{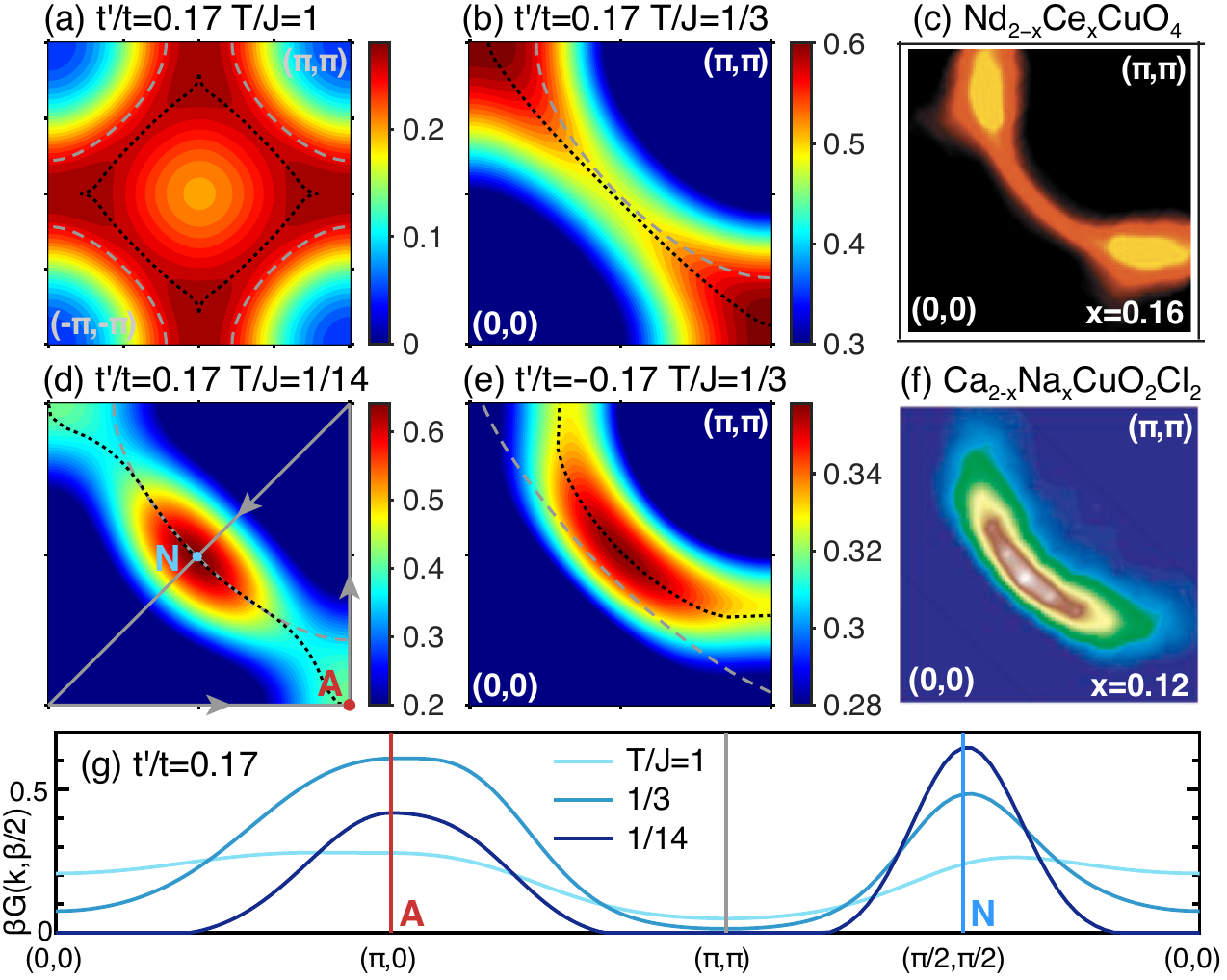}
\caption{(a,b,d) Results of $\beta G(\mathbf{k},\beta/2)$ for $T/J=1$, 
$1/3$, and $1/14$, calculated for $t'/t=0.17$ and doping ratio $\delta 
\simeq 0.12$, compared to experiments on (c) electron- and (f) 
hole-doped cuprates. The dotted black lines denote the ridges, 
signifying the shape of FS, while the dashed gray lines illustrate the 
free electron FS at the same doping level. For clarity, only the first quadrant 
is show in (b)-(f). (e) shows $\beta G(\mathbf{k},\beta/2)$ for $t'/t=-0.17$ case, 
with $T/J=1/3$ and $\delta \simeq 0.11$. (g) shows $\beta G(\mathbf{k},\beta/2)$ 
along the triangular path illustrated in (d), where the red and blue 
vertical lines emphasize the antinodal (A) and nodal (N) points.
}
\label{Fig:Matsubara}
\end{figure}

In Figs.~\ref{Fig:Matsubara}(a), (b), and (d) we show the $\beta G(\mathbf{k},\beta/2)$ results for 
$t'/t=0.17$ with fixed doping $\delta \simeq 0.12$, calculated on a $6\times 6$ 
square lattice with open boundaries. To facilitate comparison to electron-doped 
cuprates with $t'/t<0$~\cite{Markiewicz2005}, the data are shifted by $(\pi,\pi)$, 
as through the particle-hole transformation the system can be mapped to $t'/t=-0.17$ 
with electron density $n=1+\delta$. In Fig.~\ref{Fig:Matsubara}(a), for $T/J=1$ in 
the HT regime, we find no clear suppression in the FS spectral weights, while in Fig.~\ref{Fig:Matsubara}(b) 
the spectral weight gets suppressed near $(\pi/2,\pi/2)$ for $T/J=1/3$ in the PG regime, 
consistent with the ARPES results of a typical electron-doped superconductor 
Nd$_{2-x}$Ce$_x$CuO$_4$ [Fig.~\ref{Fig:Matsubara}(c)]~\cite{Matsui2007NCCO,
Armitage2010RMP}. Upon further reducing the temperature to $T/J=1/14$ and entering 
the SC phase, a dramatic shift in intensity is observed in Fig.~\ref{Fig:Matsubara}(d). 
Specifically, the intensities show pronounced increase near the nodal point $(\pi/2, \pi/2)$, 
while get suppressed around the antinodal points $(\pi,0)$ and $(0,\pi)$. This observation 
aligns well with the characteristics of $d_{x^2-y^2}$-wave pairing symmetry.

In the hole-doped case with $t'/t=-0.17$, we present the results for $T/J=1/3$ 
in Fig.~\ref{Fig:Matsubara}(e), again consistent with ARPES experiments [Fig.
\ref{Fig:Matsubara}(f)] conducted in the PG regime of a hole-doped cuprate 
Ca$_{2-x}$Na$_x$CuO$_2$Cl$_2$~\cite{Shen2005Science}. The intensity pattern 
remains qualitatively similar at lower temperatures, as detailed in Supplemental 
Material~\cite{Supplementary}, contrasting with the results observed in the SC phase 
with $t'/t>0$.

% ======= Fig. 4 ======== %

\textit{Discussion and outlook.---}
Recent studies have made great progresses in understanding the ground state 
of the $t$-$t'$-$J$ model, where robust $d$-wave SC is found for $t'/t>0$
\cite{HCJiang2021tJ,Gong2021tJ,Jiang2021PNAS,Jiang2023tJ8leg}. Here we make 
a significant step further by obtaining the finite-$T$ phase diagram of this model, 
on up to width-6 cylinder. In particular, for $t'/t=0.17>0$ we find a domelike 
SC regime, with the characteristic temperature $T_c^* \sim 0.15 J$, i.e., $\sim 0.05 t$, 
near optimal doping. The estimated transition temperature is high ($\gtrsim 100$~K), 
given $t =$ 0.3-0.5~eV for typical cuprates~\cite{Gull2013SCPG,Motoaki2019PRB}. 
Above the dome and below a higher temperature scale $T^*$, we find PG-like 
behaviors with suppressed magnetic susceptibility and a FS with node-antinode 
features. On the other hand, for $t'/t=-0.17<0$, we find a very different scenario with 
nearly negligible SC correlations but prominent CDW correlations. Moreover, the partially 
suppressed spectral weights on the FS for both $t'/t= 0.17$ and $-0.17$ cases align 
with those observed in electron- and hole-doped cuprates, respectively. 

Finite-temperature calculations have alleviated finite-size effects, at least in the high- 
to intermediate-temperature regime, where the conclusion may hold for the 2D limit. 
For $t'/t>0$, we have seen qualitative consistency between the simulated $W=4$ and 
$W=6$ phase diagrams~\cite{Supplementary}, where a robust high-$T_c$ SC regime 
is revealed; for $t'/t<0$, on the other hand, recently there are intensive studies on the 
ground-state properties, yet the existence of SC order remains inconclusive
\cite{Jiang2021PNAS,Xu2024Science,Lu2023SignPRB,Jiang2024Hubbard6leg,Lu2024tJ8leg,
Chen2023tJPDW}. On the width-6 cylinder, recent DMRG studies find an absence 
of SC but have identified the CDW order for typical doping levels of $\delta \approx 1/12$ 
to $1/8$~\cite{Jiang2021PNAS,Jiang2024Hubbard6leg,Lu2023SignPRB,Lu2024tJ8leg}. 
In our finite-temperature study, we find the pairing correlations are indeed small  
[Fig.~\ref{Fig:PhDiag}(b)], while the CDW fluctuations become prominent as temperature
decreases, consistent with previous DMRG findings. 

Moreover, possible quasi-long-range SC correlations has been reported on the $W=8$ cylinder 
for $t'/t<0$ and $\delta=1/8$~\cite{Lu2024tJ8leg}, parameters pertinent to hole-doped cuprates. 
However, the pairing correlations are clearly weaker (by an order of magnitude) compared to those 
for $t'/t>0$. In a different work on the $t$-$t'$-Hubbard model, coexistence of SC order and partially filled 
stripe order was obtained by employing twisted-averaged boundary conditions and sampling over 
low-lying states~\cite{Xu2024Science}. In contrast, we present a thermal average of low-energy eigenstates, characteristics 
of a phase rather than an individual state. Our phase diagrams shown in Fig.~\ref{Fig:PhDiag} are thus highly 
indicative for wider cylinders and impose significant constraints of the potential SC phase there. 
To address the difference between current simulations and cuprate experiments, it may be 
necessary to introduce additional terms, such as considering a three-band model or density-assisted 
hopping terms~\cite{Jiang2023ThreeBand}, etc. Besides, there are rapid progresses in the cold-atom 
quantum simulations of Hubbard/$t$-$J$ models~\cite{Mazurenko2017,Chiu2019,Bloch2022HubbardLadder,
Bloch2023tJNautre}, and our results provide fresh perspectives on these experiments at finite temperature.

%\begin{acknowledgments}
\textit{Acknowledgments}.--- W.L. and D.-W.Q are indebted to Xin Lu, Bin-Bin Chen, 
Tao Shi, Zi-Xiang Li, Wei Wu, and Lei Wang for helpful discussions. This work was 
supported by the National Natural Science Foundation of China (Grant Nos. 12222412, 
11974036, 11834014, 12047503, 12174386, 12274014, and 11874115), National Key R\&D 
Program of China (Grant No.~2018YFA0305800), Strategic Priority Research Program 
of CAS (Grant No.~XDB28000000), the Innovation Program for Quantum Science and 
Technology (under Grant Nos.~2021ZD0301800 and 2021ZD0301900), and CAS Project 
for Young Scientists in Basic Research (Grant No.~YSBR-057). We thank the HPC-ITP 
for the technical support and generous allocation of CPU time.
%\end{acknowledgments}

\bibliography{tJRefs}

%======================================
%========== Supplementary =============
%======================================

\newpage
\clearpage
\onecolumngrid

\begin{center}
{\large Supplemental Material for} $\,$
\\
\textbf{\large{Phase Diagram, $d$-Wave Superconductivity, and Pseudogap 
of the $t$-$t'$-$J$ Model at Finite Temperature}}
\end{center}

\renewcommand\thefigure{S\arabic{figure}}
\renewcommand\theequation{S\arabic{equation}}

\setcounter{section}{0}
\setcounter{figure}{0}
\setcounter{equation}{0}
\setcounter{table}{0}

\section{Convergence of the free energy versus bond dimenions}
\begin{figure}[bp]
\includegraphics[width=0.85\linewidth]{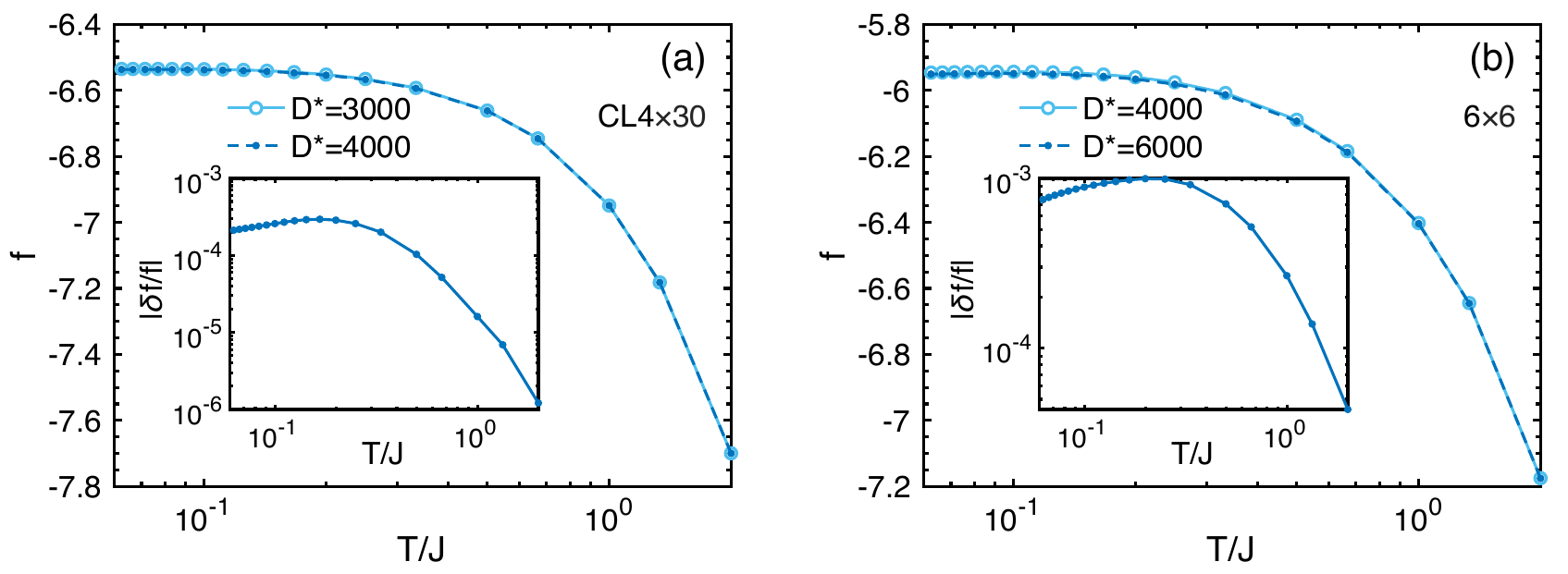}
\caption{Convergence of the free energy per site vs. $D^*$. 
(a) shows the results of $4\times 30$ cylinder ($W=4, L=30, t'/t=0.17, \mu=5.34$) 
obtained with bond dimension $D^*=3000$ (equivalent $D\simeq9000$ 
individual states) and $D^*=4000$ ($D\simeq12000$). 
(b) shows the results of $6\times 6$ open square [$W=6, L=6, t'/t=0.17, \mu=4.9$ 
with bond dimension $D^*=4000$ ($D\simeq14000$) and $D^*=6000$ 
($D\simeq 21000$)]. The insets show the relative difference $|\delta{f}/f|$
between different bond dimensions, which indicate a very good convergence 
in the free energy.
}
\label{FigS:FeConvrg}
\end{figure}

In Fig.~\ref{FigS:FeConvrg}(a), we present the calculated free energy per site of 
the $4\times 30$ ($W=4, L=30$) system. The results were obtained by keeping 
two different values of $D^*$, i.e., $D^*=3000$ and 
$D^*=4000$ multiplets, respectively. The relative difference of $f$ is $\lesssim 
3 \times 10^{-4}$, showing a very good convergence. 
In Fig.~\ref{FigS:FeConvrg}(b), we find the free energy $f$ results of the 
$6\times 6$ open square (with $\mu=4.9$) show relative difference within 
$10^{-3}$ for $D^*=4000$ and $D^*=6000$. 

Overall, the results presented in Fig.~\ref{FigS:FeConvrg} demonstrate the 
importance of carefully assessing the convergence of numerical results and 
choosing appropriate values of $D^*$ to ensure the accuracy and reliability of 
the calculations. By verifying the convergence of the free energy, we can also
be sure that other thermodynamic quantities, e.g., the specific heat, calculated 
from it are also accurate and reliable.

% //More pairing correlation data
\section{Superconductivity pairing correlations}

In this section, we provide more data on the superconductivity pairing 
correlations for 2-leg ladder and width-4 cylinder with $t'/t>0$. In the following, 
we study a 2-leg ladder with slightly different model parameters, 
namely, $t/J=3$, $J'/J=0.05$, and $t'=t\sqrt{J'/J}$. This larger $t'/t$
leads to enhanced superconductivity and suppressed CDW instabilities
\cite{Lu2022tJPRB}.

\subsection{Pairing correlation scalings in the 2-leg ladder}
Firstly, we provide a brief review of some analytical results for the 2-leg ladder, 
which will aid in the analysis of our numerical data. For the parameters mentioned 
above, and with doping ratio $\delta \leq \delta_c\simeq 0.34$, the ground state on 
2-leg ladder is a Luther-Emery liquid (LEL) with quasi-long range SC correlations. 
While for $\delta>\delta_c$ (except for the special doping $\delta=1/2$), the ground 
state is in the Tomonaga-Luttinger liquid (TLL) phase. At doping $\delta=1/2$, 
there exists a special SDW point where the system opens up a small charge 
gap~\cite{Lu2022tJPRB}, while the antiferromagnetic spin-spin correlations are still 
quasi-long-ranged. Bosonization analysis shows that the pairing correlation 
functions in LEL has two leading terms in the long-range scaling
\cite{Giamarchi1D,Hayward1996tJ}, i.e.,
\begin{equation}\label{Eq:PhiLEL}
\Phi_{\alpha\beta}(r) = \frac{C_0}{r^{1/(2K_\rho)}} + 
C_1\frac{\cos(2k_F r)}{r^{2K_\rho+1/(2K_\rho)}}, 
\end{equation}
where $K_\rho$ is the Luttinger parameter, $k_F$ is the Fermi momentum 
estimated from the particle density, and $C_0, C_1$ are non-universal constants. 
As $K_\rho>0$ the first uniform term dominates in LEL, regardless of the specific 
values of $C_0, C_1$, and the conclusion holds for different directions of pairings. 
On the other hand, the scaling behavior in TLL is \cite{Giamarchi1D,Hayward1996tJ}
\begin{equation}\label{Eq:PhiTLL}
\Phi_{\alpha\beta}(r) = \frac{C_2}{r^{1+1/(2K_\rho)}} + 
C_3\frac{\cos(2k_F r)}{r^{2K_\rho+1/(2K_\rho)}},
\end{equation}
where instead the second $2k_F$-oscillation term dominates,
as we always find $K_\rho<1/2$ in the 2-leg ladder~\cite{Lu2022tJPRB}.
 
\begin{figure}[!tbp]
\includegraphics[width=1\linewidth]{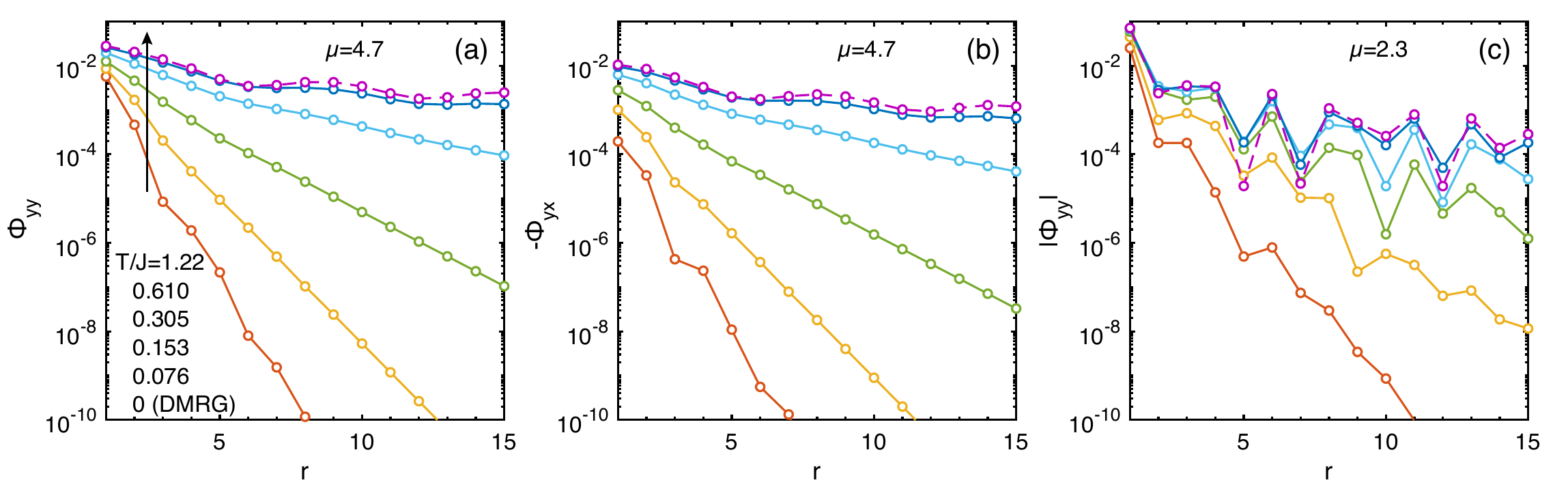}
\caption{Pairing correlation functions for 2-leg ladder at various temperatures. 
Here we show the data for an $L=32$ system. The reference point 
of correlation is at $x=L/4+1$, and the distance is denoted by $r$. (a,b) show 
respectively the $y$-$y$ and $y$-$x$ pairing correlation functions $\Phi_{yy}(r)$ 
and $\Phi_{yx}(r)$ with $\mu=4.7$ (near optimal doping, LEL phase). 
We find in (a) that $\Phi_{yy}(r)$ is always positive, while the $y$-$x$ pairing is 
negative [hence we show $-\Phi_{yx}(r)$ in (b)]. (c) shows the absolute value 
$|\Phi_{yy}|$ for $\mu=2.3$ (overdoped, TLL phase). The dashed lines are the 
ground-state DMRG results with 10 doped holes in (a,b) and 26 holes in (c).
}
\label{FigS:Pyy}
\end{figure}

In \Figs{FigS:Pyy}(a,b) we show $\Phi_{yy}(r)$ and $-\Phi_{yx}(r)$ at various 
temperatures, for the 2-leg ladder with $\mu=4.7$ (LEL ground state). According 
to the $d$-wave paring symmetry, for $\Phi_{yy}$ we have $C_0>0$ while 
for $\Phi_{yx}$ the constant is $C_0<0$, and indeed here $\Phi_{yy}>0$ 
and $\Phi_{yx}<0$ are observed. 
On the other hand, in \Fig{FigS:Pyy}(c) we show $|\Phi_{yy}(r)|$ at various 
$T$ in the TLL phase (with $\mu=2.3$), where the pairing correlations show 
clear oscillation behaviors vs.~$r$. In both phases, LEL and TLL, the pairing 
correlations obtained by thermal tensor networks converge to the 
ground-state DMRG results at sufficiently low temperatures. 

\begin{figure}[!tbp]
\includegraphics[width=0.85\linewidth]{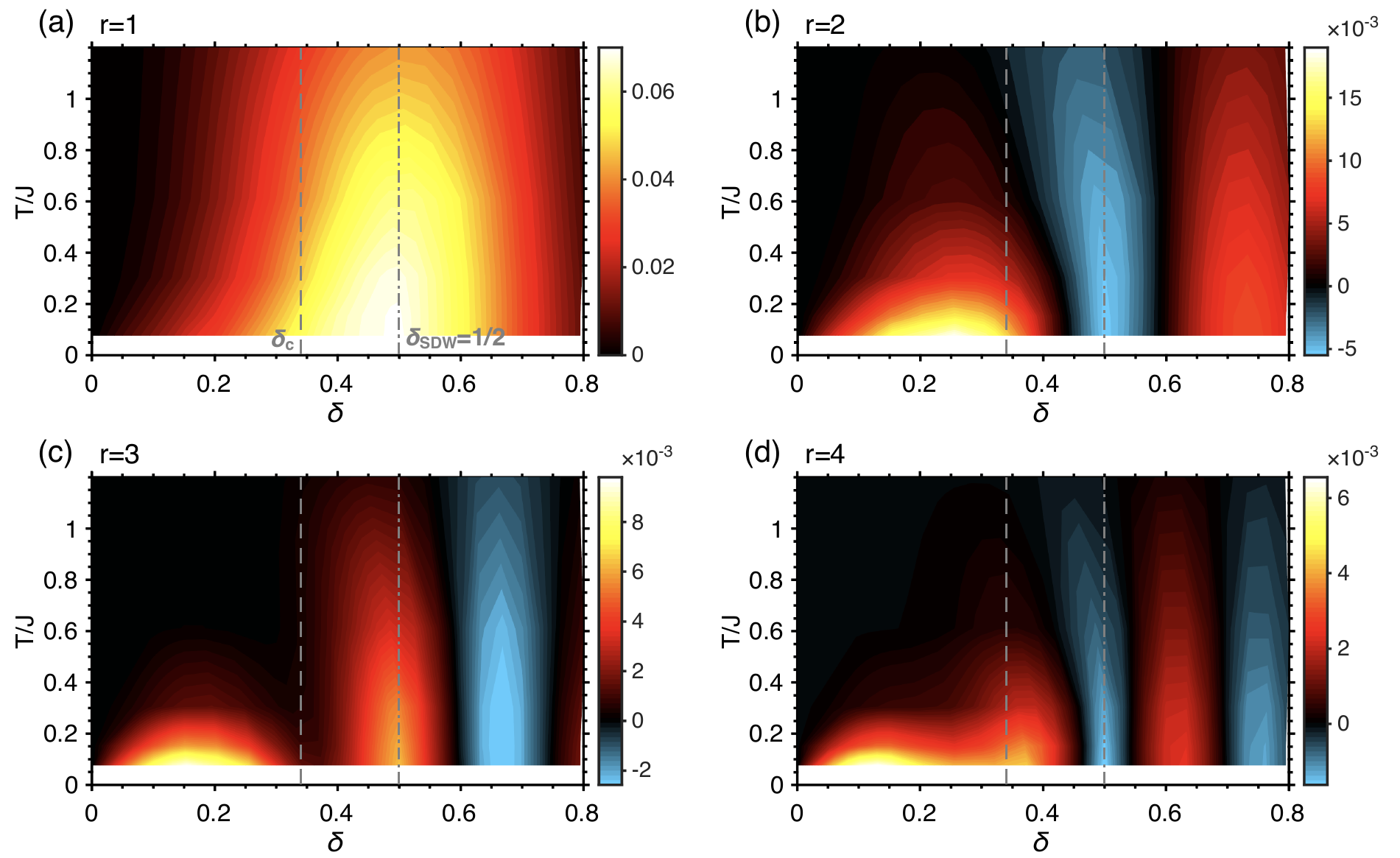}
\caption{(a-d) Contour plot of the $y$-$y$ pairing correlation $\Phi_{yy}(r)$ 
($r=1$-$4$) on a 2-leg ladder, as a function of doping $\delta$ and 
temperature $T$. The dashed lines denote the LEL-TLL critical point 
$\delta_c\sim 0.34$ in ground state, and the dotted dashed 
lines denote the SDW point at $\delta_\text{SDW} = 1/2$. 
}
\label{FigS:PyyContour}
\end{figure}

\begin{figure}[!tbp]
\includegraphics[width=0.85\linewidth]{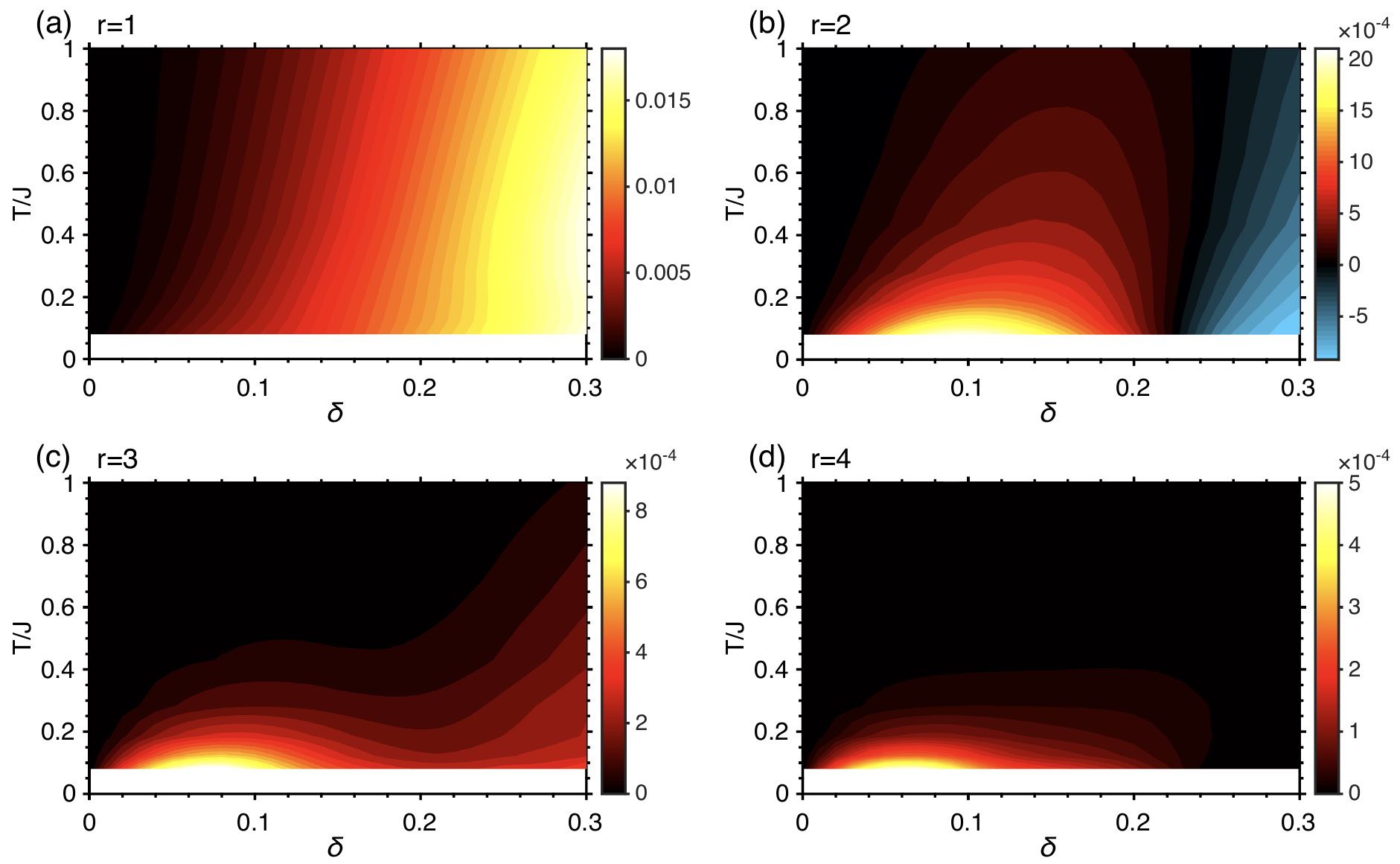}
\caption{(a-d) Contour plot of the pairing correlation $\Phi_{yy}(r)$ 
($r=1$-$4$) on the width-4 cylinder, as a function of doping $\delta$ and 
temperature $T$. 
}
\label{FigS:PyyContourW4}
\end{figure}

\begin{figure}[!tbp]
\includegraphics[width=0.5\linewidth]{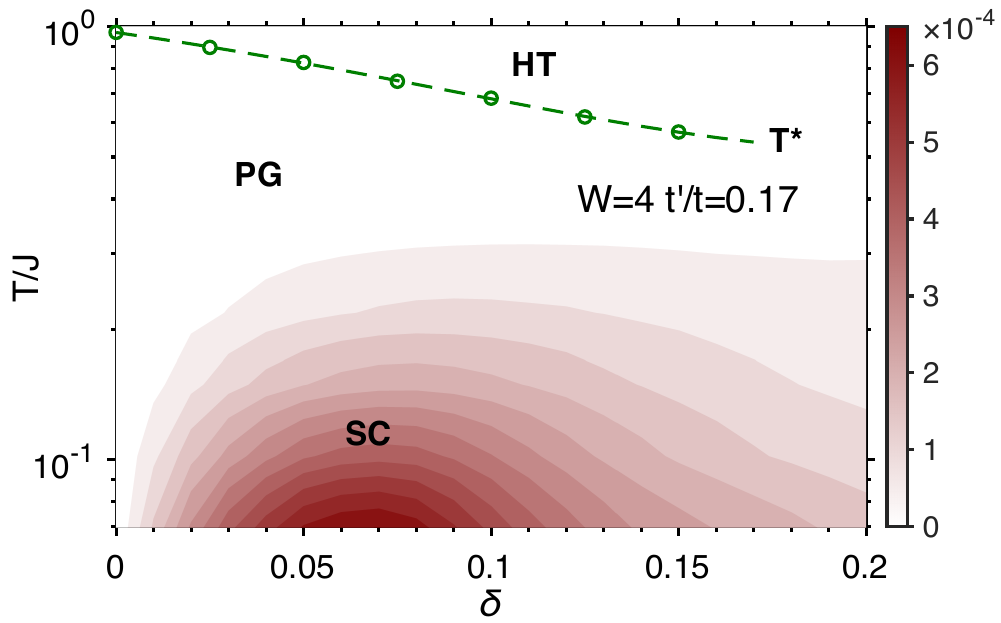}
\caption{Phase diagram for $t'/t=0.17$ on a $W=4$ cylinder. 
The red contour map shows the pairing correlations $\Phi_{yy}(r)$ averaged 
over $r=3$-$5$, which exhibits a domelike fluctuating SC regime. The dashed 
green line is the peak temperature of magnetic susceptibility $\chi_m$, 
which separate the PG and HT regimes.
}
\label{FigS:PyyW4}
\end{figure}

% ======== Contour plot ========== %
\subsection{Pairing correlation contours for the width-2 and 4 systems}

In the main text, we have shown the superconducting dome by the contour of 
pairing correlation $\Phi_{yy}(r)$ averaged over the distances $r=3$-$5$. Here 
we show that such a dome structure can also be observed in $\Phi_{yy}(r)$ at 
various fixed distance $r$. For the 2-leg ladder, Figs.~\ref{FigS:PyyContour}(a-d) 
shows the contour of $\Phi_{yy}(r)$ as a functions of $\delta$ and $T$ for $r=1\text{-}4$ 
and a wide doping range $0<\delta<0.8$. 
We find that except for the nearest-neighbor $r=1$, there is always a dome-shape
regime with prominent and positive $\Phi_{yy}$ correlations. The specific 
boundaries on the right side of the dome can nevertheless change slightly 
for different distances $r$, which is possibly influenced by the nearby TLL phase. 

For $\delta>\delta_c$, $\Phi_{yy}$ does not vanish but exhibits a vertical 
bar-like structure with alternating signs as $\delta$ changes. This observation can 
be well explained with the dominant $2k_F$-oscillation in \Eq{Eq:PhiTLL}, 
where $k_F = \pi n = \pi(1-\delta)$. As expected, in \Figs{FigS:PyyContour} 
we find $\Phi_{yy}$ oscillates like $-\cos(2\pi r \delta)$ for a fixed $T$ and $r$. 
The special SDW point $\delta=1/2$ can also be seen in the finite-temperature
calculations, where the oscillation is just a $\pi$-shift, $(-1)^{r-1}$, from panel (a) 
to (d). The amplitudes of pairing correlations at $\delta=1/2$ decay exponentially 
even in the ground state, due to the presence of a finite charge gap. Unlike the 
SC dome for $\delta<\delta_c$, these bars extent to quite high temperature $T/J 
\approx 1$. This indicates the paring correlation described in Eq.~(\ref{Eq:PhiTLL}), 
particularly the oscillating term, has already been established at relatively high 
temperature in the TLL phase, much higher than the SC correlation in the 
superconductive LEL phase. However, from Fig.~\ref{FigS:PyyContour} we see
that the paring correlations in the TLL phase decay very rapidly as $r$ increases,
while they are much more robust against increasing distance $r$ in the LEL 
regime under the dome.

For the $W=4$ cylinder, in \Fig{FigS:PyyContourW4} we find a slightly different 
scenario. Although the pairing correlations also exhibit a sign change as doping 
ratio $\delta \gtrsim 0.25$, the $\Phi_{yy}(r)$ correlations in the highly doped 
regime decay much more rapidly with $r$. For $r \geq 4$ we can hardly see the 
pairing correlation in the overdoped regime. The bright dome regime with prominent 
pairing correlations, on the other hand, is persistent and the overall shape changes 
only slightly as $r$ increases. In Fig.~\ref{FigS:PyyW4} we show the $\Phi_{yy}(r)$ 
averaged over $r=3$-$5$, like Fig.~1(a) in the main text. We find the domelike 
fluctuating SC regime is apparent. Compared with the $W=6$ data in Fig.~1(a), 
the SC dome broadens from $W=4$ to $W=6$. In Fig.~\ref{FigS:PyyW4} we also show 
the peak temperature $T^*$ of magnetic susceptibility, which decreases with increasing 
doping. Therefore, the overall phase diagram for $t'/t=0.17>0$ is qualitatively unchanged 
from $W=4$ to $W=6$. On the other hand, for $t'/t=-0.17<0$ the ground state on $W=4$ 
cylinder is the special plaquette $d$-wave state~\cite{Chung2020Plaquette}, and therefore 
it is not directly comparable with the $W=6$ data in this case.

\begin{figure}[!tbp]
\includegraphics[width=0.45\linewidth]{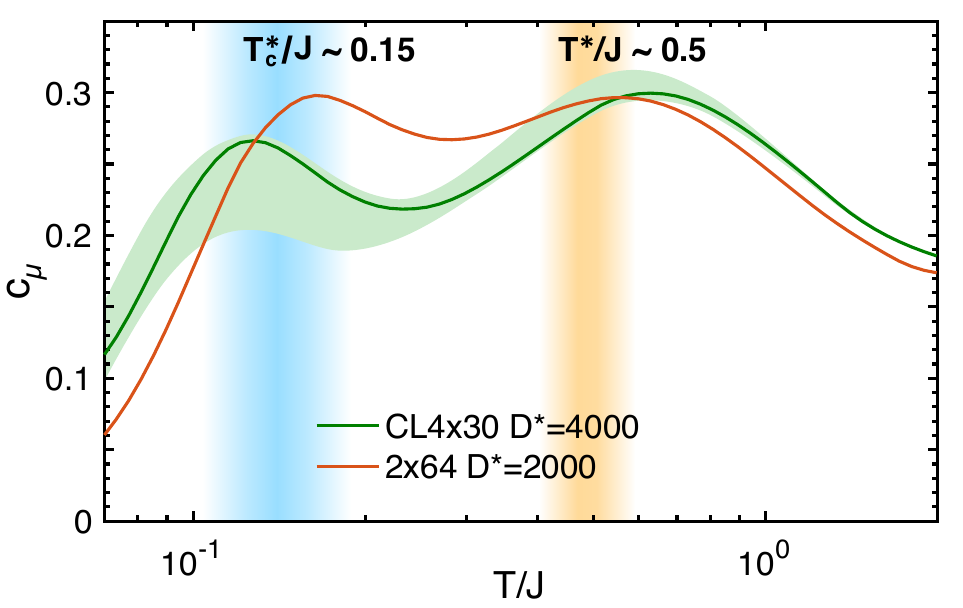}
\caption{Specific heat results calculated on a $2\times 64$ ladder with $\mu=4.7$ (and $\delta 
\simeq 0.15$) and a $4\times 30$ cylinder with $\mu=5.34$ (and $0.1 \lesssim \delta \lesssim 
0.15$). The green shadow indicates the uncertainty of the $W=4$ data, which is estimated 
from the results with different bond dimensions.
}
\label{FigS:Cmu}
\end{figure}

\subsection{Thermodynamic signature of the SC fluctuations}
The strong pairing fluctuations within the SC dome leave distinct features in the thermodynamic 
quantities. In Fig.~\ref{FigS:Cmu}, we present the constant-$\mu$ specific heat $c_\mu$ for both 
2-leg ladders and width-4 cylinders. These results indicate that $c_\mu$ exhibits a peak near the 
SC temperature $T_c^*$, which is associated with the entropy release due to the establishment 
of SC pairing correlations.

% ======= Susceptibility ====== %
\section{Magnetic susceptibility}

\begin{figure}[!tbp]
\includegraphics[width=1\linewidth]{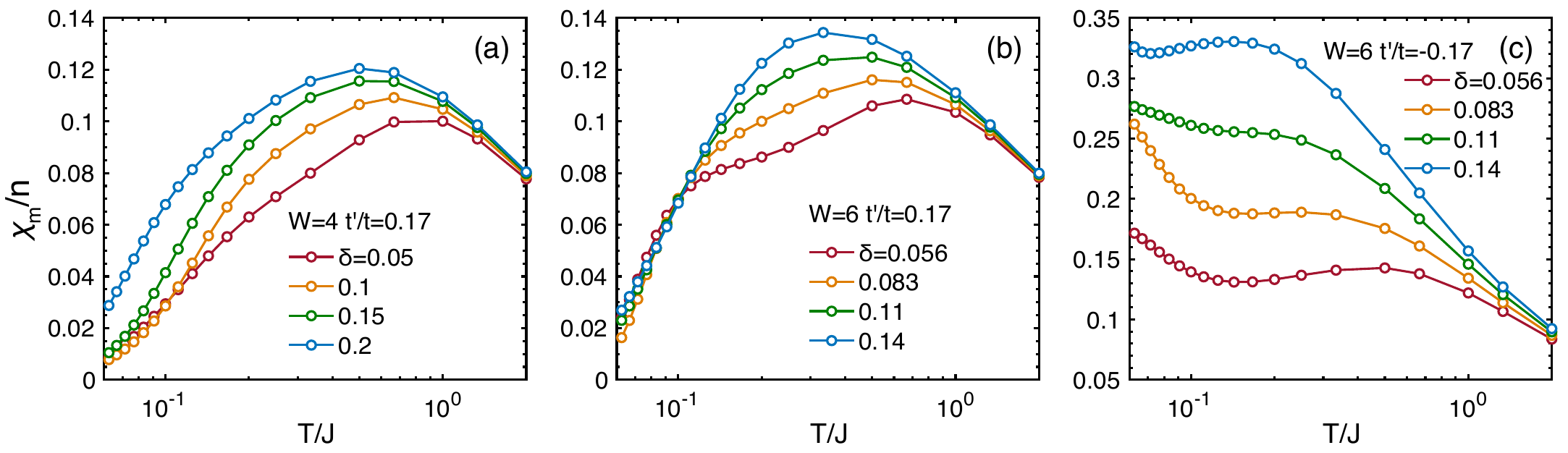}
\caption{Magnetic susceptibility $\chi_m$ as a function of $T$ for various 
doping $\delta$. For better visibility of the lines, here we show $\chi_m/n$ 
normalized by electron density $n=1-\delta$. (a) shows the results for
$W \times L=4\times 20$ cylinder with $t'/t=0.17$. (b) and (c) show the $6\times 12$ cylinder 
results for $t'/t=0.17$ and $t'/t=-0.17$ respectively. The bond dimension 
retained is $D=3000$ for $W=4$ and $D=6000$ for $W=6$.
}
\label{FigS:Chi}
\end{figure}

In Fig.~1 of the main text, we have shown the peak temperature $T^*$ of the magnetic 
susceptibility $\chi_m$, and plot $T^*$ versus doping ratio $\delta$ in the phase diagram. 
Here we provide more supporting data of $\chi_m$ with different system sizes and hopping 
ratios $t'/t$. In Fig.~\ref{FigS:Chi} we show $\chi_m$ as a function of $T$ for each 
(approximately) fixed $\delta$, by tuning the chemical potential $\mu$. In panels (a) and (b), 
we find $\chi_m$ has a peak for $t'/t=0.17$ on both $W=4$ and $W=6$ cylinders, 
below which it decreases towards zero with $T$. As the doping $\delta$ increases, 
$\chi_m$ gets enhanced and the peak is pushed to lower temperature. On the other hand, 
as shown in Fig.~\ref{FigS:Chi}(c), the values of $\chi_m$ for $t'/t=-0.17$ is much greater 
than those for $t'/t=0.17$ shown in Fig.~\ref{FigS:Chi}(b), with the same doping level. 
In this case, although the $\chi_m$ curve also exhibits a hump at $T^*$, it does not decrease
monotonically and converge towards zero below $T^*$, but increases again with $T$ at 
lower temperatures. This indicates that for $t'/t=-0.17<0$ the spin fluctuations are stronger 
in the stripe phase, while the spin-gapped superconductivity is not favored.

\section{Spin correlations}

\begin{figure}[!tbp]
\includegraphics[width=1\linewidth]{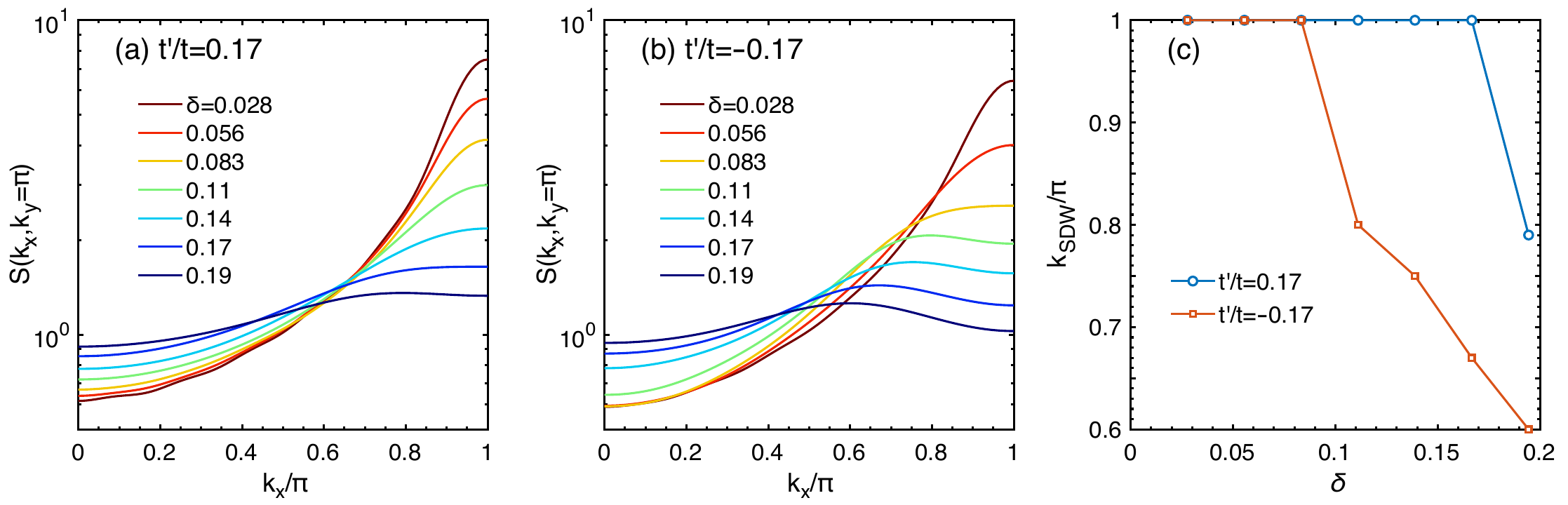}
\caption{(a) and (b) show the spin structure factor $S(k_x,k_y=\pi)$ as a 
function of $k_x$ for various doping $\delta$, for $t'/t=0.17$ and $t'/t=-0.17$ 
cases, respectively. The data are calculated on $W=6$ cylinders at a low 
temperature $T/J=1/16$. (c) shows the peak location $k_x = k_\textrm{SDW}$ of 
the spin structure factor $S(k_x,k_y=\pi)$ as a function of doping $\delta$, 
which indicates the wave vector of dominant spin correlations.
}
\label{FigS:SFSS}
\end{figure}

In Fig.~1 of the main text, we presented a contour plot of the spin structure factor $S(\pi,\pi)$ 
to illustrate the fluctuating AFM regime. Here in Fig.~\ref{FigS:SFSS}, we provide more details of 
the spin structure factor $S(k_x, k_y)$ as a function of $k_x$ (with fixed $k_y=\pi$) for various doping 
levels $\delta$, specifically for $t'/t=0.17$ in (a) and $t'/t=-0.17$ in (b). These data were calculated 
on $W=6$ cylinders at a low temperature $T/J=1/16$. For both $t'/t=0.17$ and $-0.17$ cases, 
we observe that the predominant spin correlation at low doping occurs at $(\pi,\pi)$, indicative of 
AFM behavior. As the doping level increases, the $(\pi,\pi)$ AFM peak gets suppressed for both cases,
and the peak location deviates from $k_x = \pi$ upon exceeding a specific doping level threshold.
The suppression of AFM peak is stronger for the $t'/t=-0.17$ case as compared to that of $t'/t=0.17$ 
case. In Fig.~\ref{FigS:SFSS}(c), we illustrate the wave vector $k_\textrm{SDW}$ associated with 
the dominant spin correlations. For $t'/t=0.17$, $k_\textrm{SDW}$ remains at $\pi$ until the doping 
level becomes greater than $\delta \simeq 0.17$. In contrast, for $t'/t=-0.17$, it stays at $\pi$ only 
up to $\delta \simeq 0.1$. These findings suggest that the AFM correlations are more robust in the 
$t'/t=0.17$ case. This is consistent with experiments on cuprates where the AFM regime is ``larger'' 
in the electron-doped side than that in the hole-doped side.

\section{Charge density wave}

\begin{figure}[!tbp]
\includegraphics[width=1\linewidth]{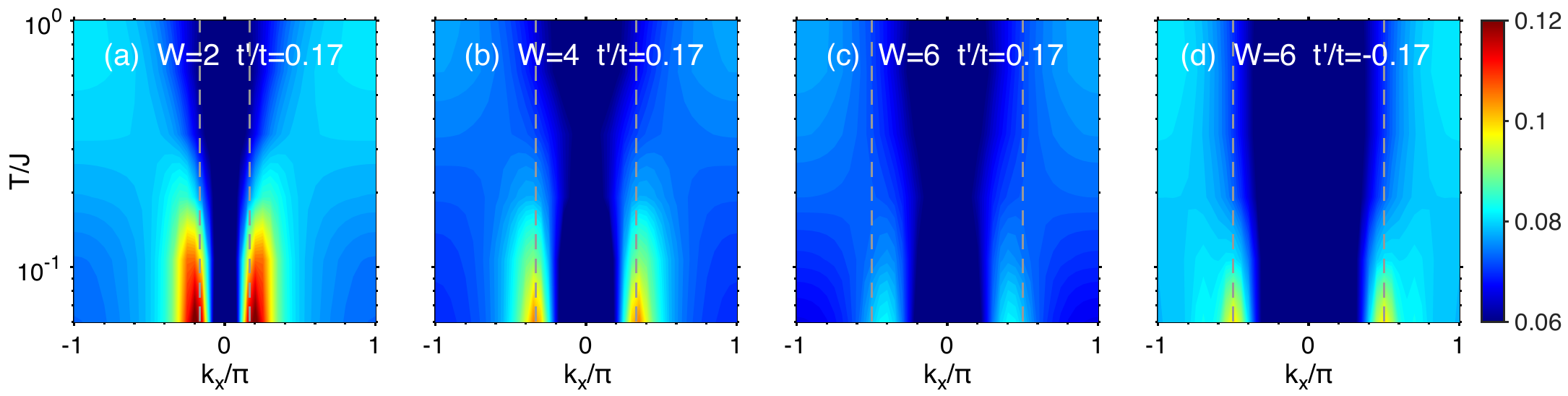}
\caption{Contour plots of charge structure factor $S_c(k_x,k_y=0)$ 
as functions of $k_x$ and $T$ for doping level $\delta\simeq 1/12$. 
(a-c) show the $t'/t=0.17$ case on $W=2,4,6$ lattices, respectively, and (d) shows 
the $t'/t=-0.17$ case on $W=6$ cylinder. The dashed gray lines indicate 
the CDW vectors $k_x\simeq\pm W\pi\delta$.
}
\label{FigS:SFnn}
\end{figure}

\begin{figure}[!tbp]
\includegraphics[width=0.75\linewidth]{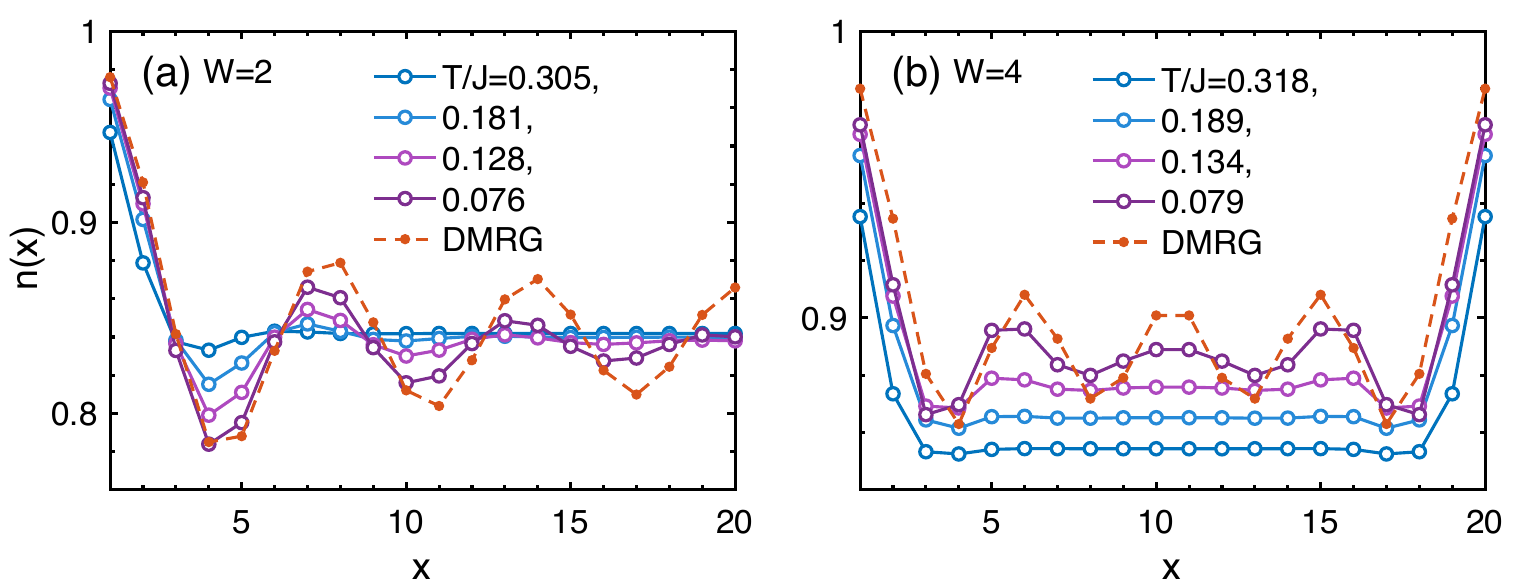}
\caption{Real-space charge density distribution $n(x)$ at various temperatures for 
(a) 2$\times$64 ladder and (b) 4$\times$20 cylinder. In (a) we plot $n(x)$ with $x \in [1,20]$ for 
the sake of clarity. The dashed red lines are the ground-state DMRG results with 20 
doped holes for the 2-leg ladder in (a) and 8 holes for the width-4 cylinder in (b). We
find the finite-temperature simulations ``shake hand" with the ground-state DMRG 
data in the zero-temperature limit.
}
\label{FigS:nx}
\end{figure}

For $t'/t=0.17>0$, we observe intertwined charge and pairing fluctuations in the dome below $T_c^*$.
In Fig.~3(a) of the main text, we have shown the charge structure factor $S_c(k_x,k_y=0)$ 
as functions of $k_x$ and $T$ for $t'/t=0.17$, calculated for $W=6$ cylinder with doping level 
$\delta\simeq 1/12$. Here, we further present the data calculated for $W=2$ ladder in 
Fig.~\ref{FigS:SFnn}(a) and $W=4$ cylinder in (b). The $W=6$ data are also shown in (c) for comparison. 
Similar to the $W=6$ data, here we find that for $W=2,4$ the CDW peaks also emerge at $k_x\simeq\pm W\pi\delta$ 
for $T\lesssim T_c^*$. However, the intensity of the peaks decrease monotonically from $W=2$ to $W=6$. 
In Fig.~\ref{FigS:SFnn}(d) we also show the $t'/t=-0.17$ data calculated for $W=6$ cylinder. Comparing 
(c) and (d), we find the CDW signature for $t'/t=-0.17$ is more prominent than the $t'/t=0.17$ case at 
the same doping level $\delta\simeq 1/12$. 

In \Fig{FigS:nx} we show the real-space charge density distribution $n(x) = \frac{1}{W}\sum_y n(x,y)$, where 
the emergence of CDW modulations can be observed. As there is a quasi-long-range order present in 
the ground state, the charge density oscillates strongly in the presence of open boundaries, i.e., there 
exists prominent Friedel oscillations that penetrates deeply into the bulk. In \Fig{FigS:nx} we show $n(x)$
for the 2-leg ladder (a) and width-4 cylinder (b) at various temperatures. As the system cools down, below 
$T_c^*/J\simeq 0.15$ the system leaves the PG regime and enter the SC dome. We find the charge distribution 
oscillations appear also below $T_c^*$ and approach the ground-state modulations in the low temperature limit.

\section{temperature evolution of spin and charge stripes}

\begin{figure}[!tbp]
\includegraphics[width=0.85\linewidth]{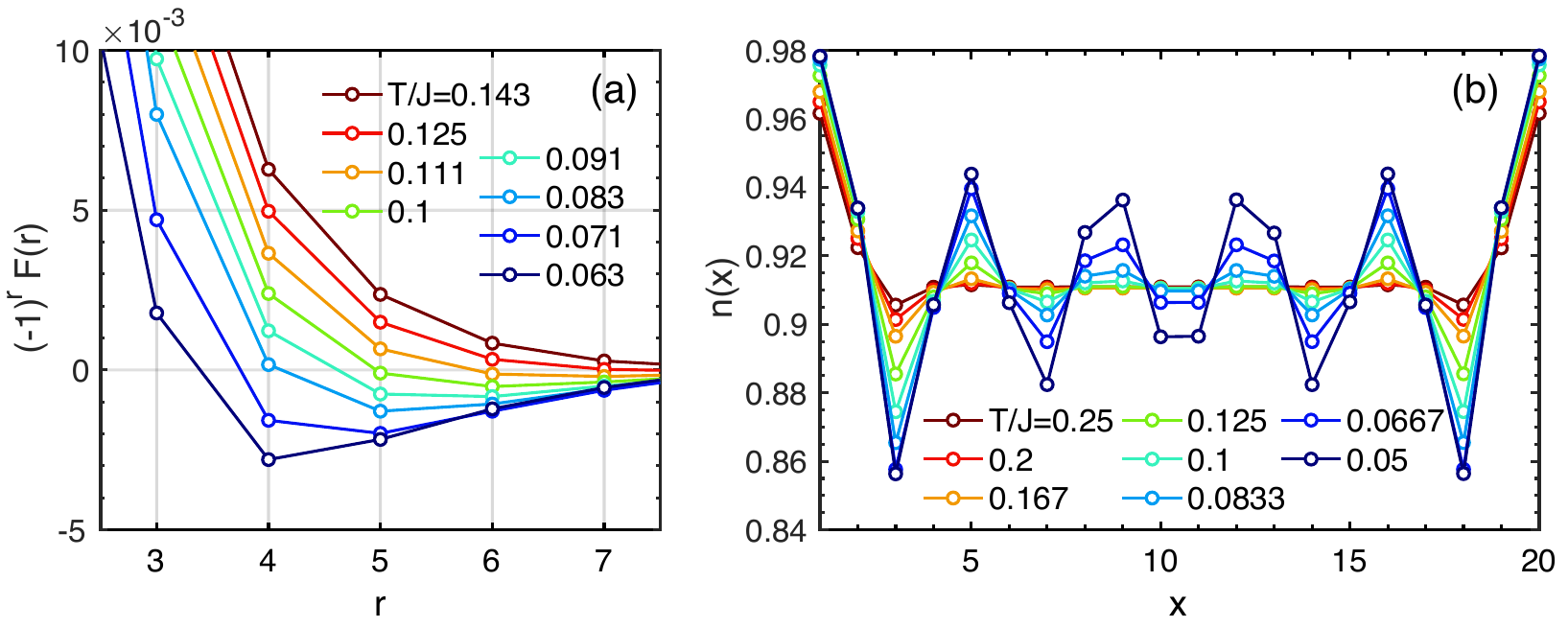}
\caption{Spin correlation and charge density distribution in real space for a 6$\times$20 cylinder system with 
$t'/t=-0.17$ and $\delta=1/12$. (a) A sign change occurs for the staggered spin correlations $(-1)^rF(r)$ along 
the cylinder (i.e., $\pi$-phase shift) below the spin-stripe onset temperature of $T^*_{\pi}/J \simeq 0.111$. 
(b) Charge density distribution $n(x)$ is shown for different temperatures, and the charge stripe (i.e., spatial 
modulation) onset temperature is estimated as $T^*_\textrm{CDW}/J \simeq 0.0833$.
}
\label{FigS:stripe}
\end{figure}

For $t'/t = -0.17 < 0$, we find that both the SDW and CDW correlations develop as temperature lowers, 
which indicates the existence of a stripe order. We show the results on a 6$\times$20 cylinder with $t'/t=-0.17$ 
and $\delta=1/12$ in Fig.~\ref{FigS:stripe}. In Fig.~\ref{FigS:stripe}(a), we show the spin correlations in real 
space (along the cylinder), and find the $\pi$-phase shift occurs for $T^*_{\pi}/J \simeq 0.111$.
As shown in Fig.~\ref{FigS:stripe}(b), we find the charge density distribution $n(x)$ exhibits 
clear spatial modulations with the wave vector $k_\textrm{CDW}=6\delta\pi=\pi/2$ below 
$T^*_\textrm{CDW}/J \simeq 0.0833$. Therefore, we conclude that the CDW onset temperature 
$T^*_\textrm{CDW}$ is slightly below the spin stripe temperature $T^*_{\pi}$. This observation is 
consistent with the scenario described in Ref.~\cite{Xiao2023OrdersPRX}, where the charge order is proposed 
to be driven by the development of spin stripe correlations. Note the stripes for $t'/t<0$ case is neither filled nor 
half-filled in our $W=6$ case, which is distinct from Ref.~\cite{Xiao2023OrdersPRX} where only filled stripes 
are observed for the $t'=0$ case. 

\section{Spectral density}

\begin{figure}[tbp]
\includegraphics[width=0.7\linewidth]{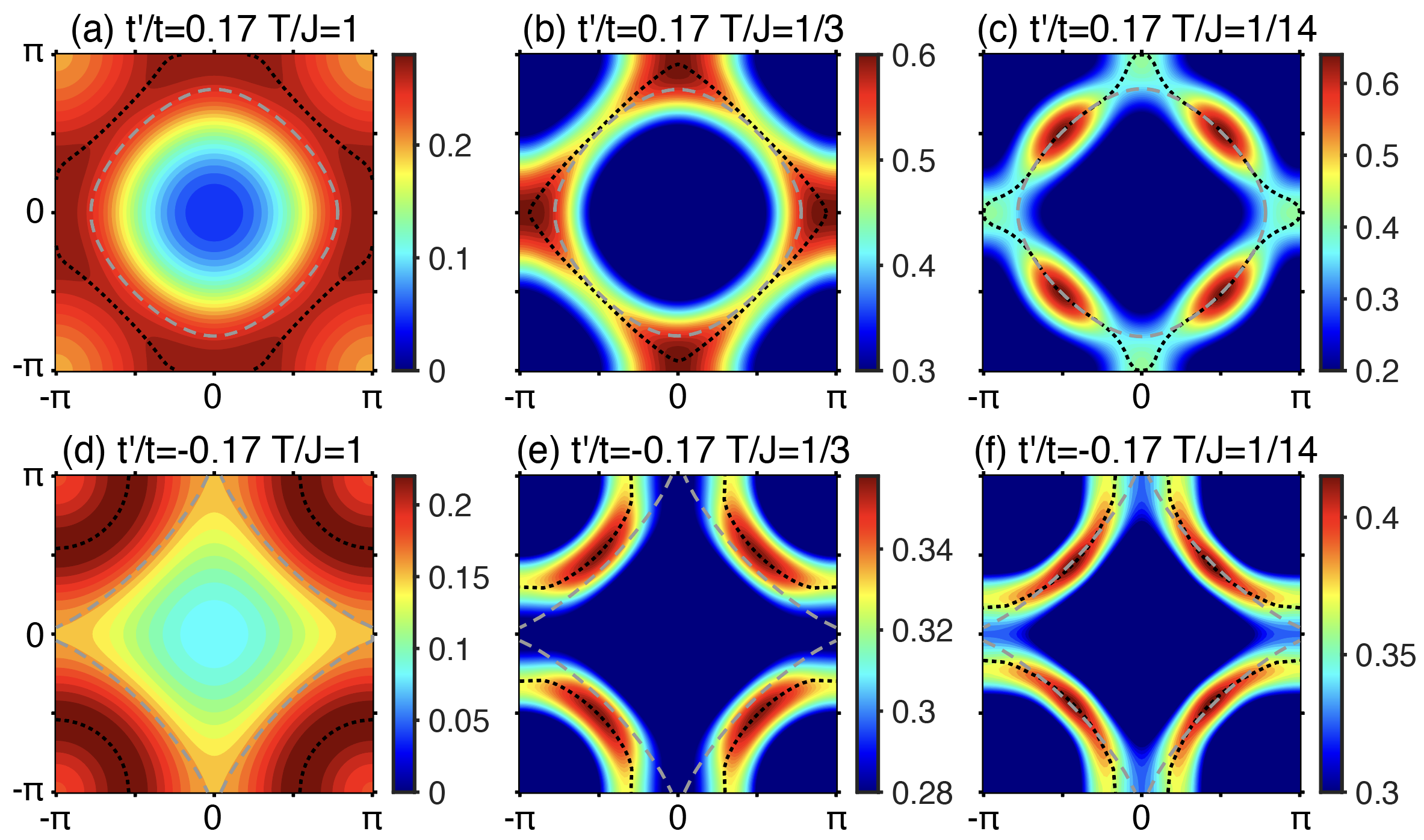}
\caption{Contour plots of $\beta G(\mathbf{k},\beta/2)$, calculated on a $6\times 6$ square 
lattice with open boundary conditions. (a-c) show the results for $t'/t=0.17$, at three
different temperatures $T/J=1$, $1/3$ and $1/14$, and with doping $\delta\simeq 0.12$. 
The data have been shown in Fig.~4(a,b,d) of the main text, where they 
have been shifted by $(\pi,\pi)$. The black dotted lines represent the ridges of $\beta G(\mathbf{k},\beta/2)$ 
and indicate the Fermi surface. The grey dashed lines represent the Fermi surface of 
the free electron system with the same hopping amplitude and doping level. 
(d-f) show the results for $t'/t=-0.17$, with temperatures $T/J=1$, $1/3$ and $1/14$, 
with doping $\delta\simeq 0.11$.
}
\label{FigS:Matsubara}
\end{figure}

In the main text, Figs.~4(a,b,d) display the temperature evolution of Matsubara Green's 
function $\beta G(\mathbf{k},\beta/2)$ for $t'/t=0.17$. In contrast, Fig.~\ref{FigS:Matsubara}(a-c) 
herein presents the same data without a $(\pi,\pi)$ shift, revealing that the Fermi surfaces 
in both the PG and SC regimes exhibit electron-like characteristics and enclose the origin
$(0,0)$, instead of $(\pi,\pi)$ after the particle-hole transformation. Despite this, the 
reduced spectral density around $(\pi/2,\pi/2)$ in the PG and the distinctive nodal structure 
associated with $d_{x^2-y^2}$-wave pairing in the SC regime remain evident.

In Fig.~\ref{FigS:Matsubara}(d-f), we present the temperature evolution of 
$\beta G(\mathbf{k},\beta/2)$ for the $t'/t=-0.17$ case. We observe that the Fermi 
surface consistently exhibits hole-like characteristics, enclosing the $(\pi,\pi)$ point. 
At high temperature $T/J=1$, the spectral density on the Fermi surface is featureless. 
However, as the temperature decreases to $T/J=1/3$, the spectral densities remain bright 
near $(\pi/2,\pi/2)$ and gets suppressed at around $(\pi,\pi)$. In contrast to the $t'/t=0.17$ case, 
further decrease in temperature does not result in a significant change in the topology of the 
intensity pattern --- it remains prominent at around $(\pi/2,\pi/2)$.

\end{document}